\documentclass{aa}
\usepackage{times}
\usepackage{epsfig}

\begin{document}

   \thesaurus{03
              (03.13.2;
               11.09.3;  
               11.17.1;  
               11.17.3;  
               11.16.1; 
               11.17.4 Q0151+048)} 
   \title{The sources of extended continuum emission towards Q0151+048A : 
	  The host galaxy and the Damped Ly$\alpha$ Absorber
             \thanks{Based on observations made with the Nordic Optical 
	      Telescope,
              operated on the island of La Palma jointly by Denmark, Finland,
              Iceland, Norway, and Sweden, in the Spanish Observatorio del
              Roque de los Muchachos of the Instituto de Astrofisica de
              Canarias.}}

   \author{J.U. Fynbo
          \inst{1,2}
          \and
           I. Burud
          \inst{3}
          \and
          Palle M\o ller
          \inst{2}
          }

   \offprints{J.U. Fynbo}
   \mail{jfynbo@obs.aau.dk}

   \institute{
           Institute of Physics and Astronomy,
           University of \AA rhus, DK-8000 \AA rhus C, Denmark
        \and
           European Southern Observatory, Karl-Schwarzschild-Stra\ss e 2, 
	   D-85748, Garching by M\"unchen, Germany 
        \and
           Institut d'Astrophysique et de Geophysique de Li\`ege, 
           Universit\'e de Li\`ege, 
	   Avenue de Cointe 5, B-4000 Li\`ege, Belgium
             }

   \date{Received ; accepted }

\titlerunning{The host galaxy of Q0151+048}
   \maketitle

\begin{abstract}
We present deep imaging in the U, B and I bands obtained under excellent
seeing conditions of the double quasar Q0151+048A,B and of the Damped
Ly$\alpha$ (DLA) absorbing galaxy at $z_{\rm abs} = 1.9342$ named S4.

We analyse the data employing two separate and independent methods.
First we deconvolve the images using the MCS algorithm, secondly we
decompose the images via an object based iteration process where we
fit models to objects without any attempt to improve the resolution
of the data.
Our detailed analysis of the images reveals, somewhat surprisingly, that
extended objects centred on the quasars themselves are much brighter
continuum sources than the DLA galaxy.

Due to the complexity caused by the many superimposed objects, we are
unable to certify whether or not continuum emission from the DLA
galaxy is detected. Continuum emission from the extended objects
centred on the positions of the quasars is clearly seen, and the
objects are tentatively identified as the ``host galaxies'' of the
quasars. The flux of those host galaxies is of order 2--6\% of the
quasar flux, and the light profile of the brighter of the two is
clearly best fit with a de Vaucouleurs profile. We discuss two
alternative interpretations of the origin of the extended flux:
{\it i)} the early stage of a massive
elliptical galaxy in the process of forming the bulk of its stars, and
{\it ii)} quasar light scattered by dust.

\end{abstract}
%
%________________________________________________________________

\section{Introduction}

The study of the galaxy population at high redshifts has progressed
rapidly during the last decade. Through the Lyman-break
technique hundreds of normal (i.e. not dominated by active galactic
nuclei), star forming galaxies at z=2--4 have been detected and 
studied with imaging as well as spectroscopy (Steidel et al. 1996).
These so called Lyman-Break Galaxies (LBGs) have star formation 
rates (SFRs) in the range 4--55h$^{-2}$ M$_{\sun}$ yr$^{-1}$
for $\Omega$=1.0 or 20 -- 270 M$_{\sun}$ yr$^{-1}$ for $\Omega$=0.2
(Pettini et al. 1998). Also, via the study of the class of 
high column density QSO absorption lines systems known as 
Damped Ly$\alpha$ Absorbers (DLAs) a wealth of information on the 
early chemical evolution of galaxies at z=2--4 has been obtained  
(e.g. Lu et al. 1996). The DLAs are in general forming stars at a 
significantly lower rate than the LBGs (M\o ller \& Warren 1998, 
Fynbo et al. 1999). 

Independent information about the formation of the brightest galaxies 
comes from detailed studies of the stellar populations of present day 
bright cluster ellipticals. These populations seems to have formed 
early (z$>$2) in strong burst of star formation (Bower et al.
1992). Studies of the fundamental plane for elliptical and
lenticular galaxies in rich clusters at intermediate redshifts
also indicate early formation times (z$>$5 for $\Omega$=1, 
J\o rgensen et al. 1999), and the fundamental plane for field ellipticals
at similar redshifts is consistent with being the same as in clusters
(Treu et al. 1999a).  Studies of the globular cluster populations 
of faint elliptical galaxies also indicate rather early formation 
times (z$>$1), whereas for bright cluster ellipticals the globular 
cluster populations do not strongly constrain the possible formation 
scenarios (Kissler-Patig et al. 1998). Furthermore,
the presence of seemingly old stellar populations in elliptical galaxies
at z$>$1 proves that at least some elliptical galaxies formed very early 
in strong bursts of star formation (Spinrad et al. 1997, 
Treu et al. 1999b, see also Jimenez et al. 1999). For first-rank 
ellipticals star formation rates as high as 
SFR$\sim$10$^{3}$ M$_{\sun}$ yr$^{-1}$ would then be possible. 
A reason why such high star formation rates have not been detected in
galaxies at high redshift may be that these galaxies are the hosts of
powerful QSOs and hence are hidden by the light from the QSOs (e.g.
Terlevich \& Boyle 1993). Support for a connection between QSOs 
and bright elliptical galaxies comes from the fact that radio quiet QSOs
as well as radio loud QSOs and radio galaxies at z=0.1-0.3 
are hosted by galaxies for which the light profiles are best fit by
de Vaucouleurs profiles indicating that they are early stages of
massive ellipticals (McLure et al. 1999).
There is increasing evidence that QSOs at redshifts z$\approx$2 are embedded 
in extended emission that is consistent with the presence of a stellar 
population in the QSO host galaxies. In the case of radio loud QSOs 
host galaxies have been detected in the optical and
infrared by Lehnert et al. (1992) and Carballo et al. (1998), and in
the case of radio quiet QSOs host galaxies have been detected in the
optical and near infrared by Aretxaga et al. (1998a,b). There
does not seem to be any systematic differences between the
host galaxies of radio loud and radio quiet QSOs. Both populations
of host galaxies are extremely bright, R$\approx$21--22, and have 
optical-to-infrared colours in the range R-K$\approx$3--5. However,
measured polarisation of the light from some 
radio galaxies show that scattered QSO light can also contribute
significantly to the observed extended emission (e.g. Cimatti et al.
1998).

In 1996  we performed  a  narrow band   study  of the
z$_{abs}\approx$z$_{em}$ Damped Ly$\alpha$ Absorber (DLA, Wolfe et al.
1986) towards Q0151+048A using the 2.56-m Nordic Optical Telescope
(NOT)  (Fynbo et al. 1999). The main result of this study
was the detection of extended Ly$\alpha$ emission from the DLA. The
Ly$\alpha$ emission line had prior to this been detected in a spectroscopic
study of Q0151+048A (M\o ller et al. 1998), but the large extended
nature of the DLA absorber was quite unexpected. U
band data, also from the 1996 run, hinted at the existence of an
extended broad band object, but the
signal--to--noise ratio of the object was low. We have therefore
obtained deeper imaging of Q0151+048 in broad band U, B and I filters
in order to confirm or reject our tentative detection, and to measure
the extend and luminosity of the broad band source if real.

   In Sect. 2 below we describe our new observations. In Sect. 3
we describe in detail the two independent methods we have used to
search for extended objects close to the quasar. First we describe
the image-deconvolution, where we used the Magain et al.
(1998, hereafter MCS) algorithm, secondly we describe the direct
PSF subtraction, and Sect. 4 we discuss our results.

   In this paper we adopt H=100 h km s$^{-1}$ Mpc$^{-1}$,
$\Omega_m$=1.0 and $\Omega_{\Lambda}$=0 unless otherwise stated.

%__________________________________________________________________

\section{Observations and Data Reduction}
The  observations were performed during two 
observing runs in September and October 1998 with HiRAC (High Resolution 
Adaptive Camera) on the 2.56 m Nordic Optical Telescope. The CCD used was 
a 2048$^2$ back-side illuminated thinned Loral with a pixel size of 0.1082 
arcsec. 
\begin{table}[t]
\caption[]{Observations of Q0151+048, Sept 17 -- 20 and Oct 17 --
18, 1998}
\begin{flushleft}
\begin{tabular}{@{}lcccccc}
\hline
Filter &  Combined seeing   & Total integration \\
       &  (arcsec) & (sec)    \\
\hline
U    & 0.96  &  12500   \\
B    & 0.84  &  24300   \\
I    & 0.67  &  11750   \\
\hline
\end{tabular}
\label{journal}
\end{flushleft}
\end{table}
    
    All 6 nights during which the observations were performed were
photometric and with good seeing (median 0.7 arcsec FWHM in the I--band). 
Integration
times in U were 1000-1500 sec in order to avoid the total noise to be 
dominated by readout noise. In I and B the integration times were
250-300 sec and 300-500 sec respectively. Between exposures the
telescope was moved 2-4 arcsec to minimize the effects of bad pixels
and fringing. The total integration times in each filter are given in
Table~\ref{journal}.

    Also observed were several standard star sequences from Landolt
(1992) and photometric solutions were obtained for each filter. For
counts given as electrons per second we derive zero-points in the 
Landolt system of 22.77, 25.13 and 24.48 for u, B and I respectively.
All magnitudes subsequently quoted in this paper are on the AB system. 
For the U--band 
we determined the colour equation $u=U+0.17(U-B)$ relating the 
instrumental magnitude $u$ to the standard Johnson $U$. There was 
substantial scatter around the fit near $U-B=0$ of $\sim 0.05$ mag., 
which is typical for this band (e.g. Bessell 1990). The instrumental 
magnitudes $u$ were converted to AB magnitudes using the equation 
$u(AB)=u+0.58$, determined by integrating the spectrum of the star 
GD71 over the passband. Here we have retained the lower case $u$ for 
the AB magnitude indicating that the effective wavelength of the filter 
lies significantly away from the
standard value. The colour term for the I and B filters are consistent 
with zero i.e. $i=I$ and $b=B$, and we used the equations $I(AB)=I+0.43$
and $B(AB)=B-0.14$ (Fukugita et al. 1995) to put the I and B magnitudes 
onto the AB system.

   The data were bias-subtracted, and twilight sky frames were used 
to flatten the frames in the standard way.

   For the deconvolution, the individual bias subtracted and flattened 
I--band  images were divided into six groups (chronologically) and the 
frames of each group were then combined using the optimal combination 
code described  by M\o ller \& Warren (1993), which maximizes the  
signal--to--noise ratio for faint sources. These six combined images 
were used in the simultaneous deconvolution process (see  
Sect.~\ref{deconv}  below). Furthermore, all the individual I--band 
images were combined into one combined image, which was used in the PSF
subtraction described in Sect.~\ref{PSF}. In the
same  way we  divided  the  individual  bias  subtracted and flattened
B--band  frames in four groups and  calculated combined images for each
group and for all images. Finally  we combined the ten individual bias
subtracted and flattened U--band frames into one combined image. The 
details of the sky noise in the combined images are provided in 
Table~\ref{sky}.
%=====================Begin Table 2====================================
\begin{table}
 \begin{flushleft}
 \caption{Measured sky level and rms of sky surface brightness.}
 \begin{tabular}{@{}lcccccc}
  passband & sky level & rms SB\\
  \hline
           & mag. arcsec$^{-2}$ & mag. arcsec$^{-2}$ \\
  \hline
  I(AB)    & 19.4 & 27.3 \\
  B(AB)    & 22.1 & 28.8 \\
  u(AB)    & 22.2 & 27.6 \\
  \hline
 \end{tabular}
 \label{sky}
 \end{flushleft}
\end{table}

%=====================End Table 2====================================

%**********Deconvolution part***********************************

\section{Analysis}
\subsection{Deconvolution}
\label{deconv}
\subsubsection{The deconvolution method}

The images  were deconvolved using  the MCS algorithm.  This method is
based on the principle that the resolution of a deconvolved image must
be  compatible  with its sampling,   which is limited  by  the Nyquist
frequency.  The deconvolved   image is   decomposed  into a   sum   of
deconvolved point--sources plus a  background smoothed on the
length scale of the final  resolution.  The intensities and  positions
of the point--sources as well as an image  of the more extended objects
are given as output of the deconvolution procedure. Image decomposition 
allows objects blended with or even superposed on point--sources to be 
studied in some detail.

In  order to check  if  the deconvolved model is compatible with the
data, a residual map is computed. The residual map contains in each
pixel the $\chi^2$ of the fit of the model image (re--convolved  with
the PSF) to the original data of that pixel. The $\chi^2$ image is used
to determine the  appropriate weight  attributed to the smoothing of the
image of extended sources in order to avoid under- or over-fitting of
the data (see  MCS for further details). A deconvolution compatible
with  the data  should show a flat residual map with a mean value of 1
all over the image. 

The MCS  algorithm  makes  it  possible  to simultaneously  deconvolve
several frames. The advantage of this process is to derive  the
optimally  constrained   deconvolved  frame  which  is  simultaneously
compatible with  several  different images of  a   given object.  This
results   in  a  more accurate  decomposition   of  the  data than the
deconvolution of one   single combined frame.  Moreover, applying  the
algorithm to many dithered frames leads to a deconvolved image with an
improved  sampling. 

\subsubsection{Application to the data}

Simultaneous deconvolution of the U--band  data  from 1996 had  already
strongly indicated extended broad band   emission in the direction  of
Q0151+048A (see Fig.~\ref{uold}).  However, although  the shape of the
extended emission was similar to  the one found in narrow-band (Fynbo
et al. 1999), it was unclear  to which extent systematic
errors in the  determination of the PSF  influenced  the detection and
hence the  signal-to-noise ratio of  the  object was too uncertain  to
constrain its morphology and luminosity.

There are two bright  stars in the field of  Q0151+048, referred to as
psfA and  psfB (see Fynbo et al.  1999).   However, our new 
deep I--band data revealed that psfB has a faint red companion 
star at a projected distance of 0\farcs7. In the I--band it is 4.3
magnitude fainter than psfB, in the B--band it is 6.3 magnitudes fainter 
than psfB, and in the U--band it remains undetected. In the following we 
only use psfA for the determination of the PSF. 

We adopted for the deconvolved image a pixel size of 0\farcs0541 (half
of the original one),   and a final   resolution of 3 pixels  FWHM, or
0\farcs16 (the Nyquist limit is 2 pixels FWHM).

\subsubsection{Results}
 
\begin{figure}[tb]
\begin{center}
\leavevmode
\epsfxsize=7. cm
\epsffile{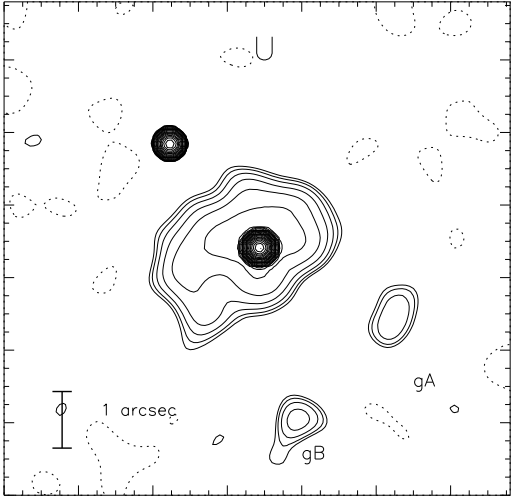}
%\vspace*{4mm}
\caption{   Contours   of the deconvolved   U--band   from  1996.   The
resolution is 0\farcs35  and  the axis are in  units  of arcsec.   The
contour levels are -9, -6, -3, 3 $\times$ 1$\sigma$ of the sky noise and 
thereafter spaced logarithmically  in  factors of 1.5, with the dotted 
contours being
negative.  The field is 12$\times$12 arcsec$^2$, North is up and
East to the left.}
\label{uold}
\end{center}
\end{figure}

\begin{figure*}
\epsfig{file=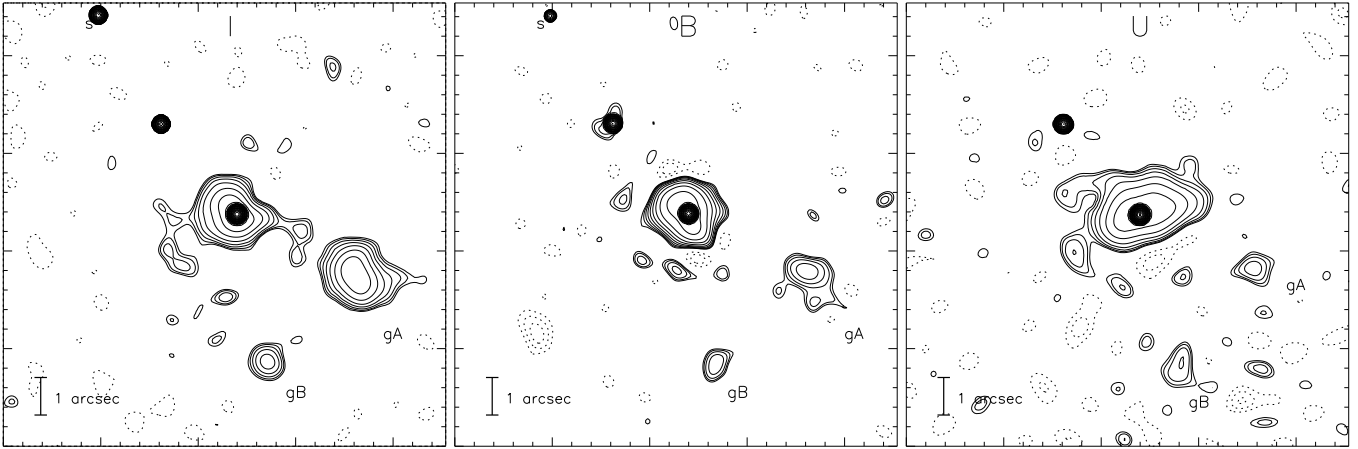,width=18cm}
\caption{Contour plots of   12$\times$12 arcsec$^2$ surrounding  the two QSOs
from the deconvolved  I--band  frame ({\it  left)}, B--band frame  {\it
(middle)} and U--band frame  {\it (right)}. The three point--sources s, qB and
qA are seen on the diagonal from the upper left corner (s) to the centre (qA).
The resolution is 0\farcs16.  North is  up and east to  the
left. Seen is the extended emission under  qA and the two galaxies gA
and gB  south  of qA.}
\label{contdec}
\end{figure*}

Fig.~\ref{contdec} shows the deconvolved images in all three bands I, B 
and U. The images show the five sources already known to be in the field, 
namely the three point--sources qA, qB and the star s, and the two faint 
galaxies gA and gB south west of qA (see Fynbo et al. 1999 for  
details). However, there is also significant extended emission under the
point--source emission from qA in all three bands. This emission have nearly
identical morphology in the I and B bands with contours centred on the 
position of qA and with a slight elongation with position angle $\sim 20^o$ 
east of 
north. The shape and intensity of the extended U--band emission under qA 
(Fig.~\ref{contdec} right panel) is consistent with the shape and intensity 
derived from the 1996 U--band data (Fig.~\ref{uold}). Since the two sets of 
U--band data have been obtained with two different instruments, the consistency 
between the two measurements makes strong systematic errors unlikely. The 
morphology of the U--band emission is significantly more extended than the 
B and I morphology, about 4.5$\times$2.2 arcsec$^2$, and has a position angle  
of about 100$^\circ$ east of north. This is very similar to that of the
Ly$\alpha$ source S4, which extends over 6$\times$3 arcsec$^2$ with
position angle 98$^\circ$ east of north.

The extended emission towards qA is $\sim$4 magnitudes fainter than that of
the point--source emission.  This high contrast makes it difficult to
determine the exact ratio between the luminosity of  the extended
source ($L_{\rm Ext}$) and that of qA ($L_{\rm QSO}$). Several
deconvolution solutions with different luminosity $L_{\rm QSO}/L_{\rm Ext}$
ratios are compatible with the  residual map constructed as described above.
Thus, there  is some degeneracy between  the plausible solutions found
by the algorithm.

In order to demarcate the range of plausible solutions, a grid of 15
deconvolved images in each band was calculated, representing 5 different
luminosity ratios $L_{\rm QSO}/L_{\rm Ext}$ and with 3 different values of
the Lagrangian smoothing parameter applied during the deconvolution (see
MCS for a description of the Lagrangian smoothing parameter). The solutions
with the highest $L_{\rm QSO}/L_{\rm Ext}$ ratios were unphysical since
they have a ring-shaped morphology, i.e. a hole at the position of the
QSO. The lowest values of $L_{\rm QSO}/L_{\rm Ext}$ were rejected 
by inspection of the residual map mentioned above. The resulting range of 
magnitudes for the extended emission is given in Table~\ref{photometry}.
The solutions shown in Fig.2 are those with the highest acceptable
values of $L_{\rm QSO}/L_{\rm Ext}$. Our conclusions concerning the
morphology of the extended emission in the three bands are, however,
unchanged for all solutions within the acceptable range.

   In the B-band there is also significant extended emission under
the PSF of the fainter neighbour quasar qB.

%=====================Begin Figure 1====================================
%=======================End Figure 1====================================

%=====================Begin Figure 2====================================
%=======================End Figure 2====================================

%=====================Begin Table 3====================================
\begin{table}[t]
\caption[]{Photometry of qA, qB and the extended sources under qA and 
qB. The magnitudes given for ExtA under Deconvolution  
is measured in a circular aperture with diameter 3.5 arcsec.  
The magnitudes for S4 and HGa under PSF-subtraction are determined 
by model fits as described in the text. The upper limits to the
magnitudes of HGb are 2$\sigma$. The magnitudes of gA and 
gB are measured in a circular aperture with diameter 1.35 arcsec.) 
}
\begin{flushleft}
\begin{tabular}{@{}lcccccc}
\hline
 &  u(AB)  & B(AB) & I(AB)  \\
\hline
Deconvolution &              &                    &                  \\
qA      & 17.81$\pm$0.02     & 17.83$\pm$0.02     & 17.46$\pm$0.02   \\
qB      & 21.00$\pm$0.02     & 21.18$\pm$0.02     & 20.69$\pm$0.02   \\
gA      & 23.85$\pm$0.09     & 23.15$\pm$0.02     & 21.77$\pm$0.02   \\
gB      & 24.02$\pm$0.09     & 24.06$\pm$0.02     & 23.18$\pm$0.03   \\
ExtA    & 21.8 -- 21.3       & 21.3 -- 20.7       & 21.8 -- 21.2     \\
NOT96   &                    &                    &                  \\
ExtA    & 21.8$\pm$0.2       &      -             &      -           \\
gA      & 24.08$\pm$0.06     &      -             & 21.87$\pm0.03$   \\
gB      & 24.18$\pm$0.07     &      -             & 23.33$\pm0.05$   \\
\hline
PSF-subtraction & & & \\
HGa      & 21.7$\pm$0.2       & 20.8$\pm$0.2      & 21.5$\pm$0.2     \\
HGb      & $>$26.0            & 25.00$\pm$0.10    & $>$25.7          \\
S4       & 23.9$\pm$0.3       & 24.1$\pm$0.3      & 23.7$\pm$0.3     \\
gA       & 24.27$\pm$0.12     & 23.51$\pm$0.03    & 22.10$\pm$0.03   \\
gB       & 24.28$\pm$0.12     & 24.53$\pm$0.06    & 23.82$\pm$0.09   \\
\hline
\end{tabular}
\label{photometry}
\end{flushleft}
\end{table}
%========================End Table 3===================================

%=====================her starter sect 3.2==============================

\subsection{Object based image decomposition}
\label{PSF}

In conclusion of the previous section: {\it i)} There is clear evidence
for extended broad band (U, B and I) emission in the vicinity of the
quasars Q0151+048A,B;
{\it ii)} the morphology of the extended object(s) is identical in B
and I but significantly different in U;
{\it iii)} the U--band morphology is more extended and similar to the
morphology of the Ly$\alpha$ emission from the DLA absorbing galaxy
(S4).

Those conclusions would suggest that the extended emission in this
field is made up of three individual components: The DLA absorbing
galaxy, the host galaxy of qA and the host galaxy of qB. The different
morphology in the four different bands would then indicate that the
objects have different spectral energy distributions (SEDs). 

In order to investigate this further we decomposed the superimposed 
images into individual objects with different SEDs.
For this image decomposition we applied the same procedure we used
for the narrow band image analysis (Fynbo et al. 1999), but here we add
more components. We consider point--sources, de Vaucouleurs profiles and
exponential profiles. The best decomposition is determined as the
minimum $\chi^2$ fit following an iterative procedure as described
below.

\subsubsection{B--band data}

The B--band image is more than a magnitude deeper than the I and
U--band images in terms of the background rms surface brightness. 
Our first step was therefore to produce optimized models of the galaxies
from the B--band data. For the decomposition we considered the
following 8 components: Three point--sources (qA, qB, s), four galaxies
to be fitted (gA, gB and the host galaxies of qA and qB, in what
follows named HGa and HGb), and one galaxy of ``frozen'' morphology
(the DLA absorbing galaxy S4). For S4 we adopted the model
determined from the narrow--band data (Fynbo et al. 1999). Note that
most of the objects do not overlap significantly, thus allowing us to 
fit them independently.

The bright star psfA was used with DAOPHOT II (Stetson 1997) to define
the PSF. We then employed the iterative $\chi^2$ minimization procedure
detailed in Fynbo et al. 1999, to decompose the image of qA into a
point--source and a de Vaucouleurs galaxy model (convolved with the
PSF). For the calculation of the $\chi^2$ we excluded a circular region
of radius 0.65 arcsec centred on qA due to the large PSF-subtraction
residuals. After ten iterations a stable solution was found. The
same procedure repeated with an exponential-disc profile instead of
the de Vaucouleurs profile resulted in a much poorer fit. A
significant positive residual, centred about 1 arcsec east of qA, was
left after this procedure. The most plausible interpretation of the
residual is that it originates from the DLA absorbing galaxy S4. As
S4 is known to extend across the position of the bright quasar,
measuring its flux from a direct aperture measurement is impossible
because of the large PSF subtraction residuals. Instead we made a grid
of models to determine the flux of S4 via minimum $\chi^2$ fitting.
For a given assumed B magnitude of S4 we first subtracted the correctly
scaled exponential--disc model as determined from
the original Ly$\alpha$ image (Fynbo et al. 1999).
For that given B magnitude of S4 we then repeated the
iterative fitting of a de Vaucouleurs profile to the remaining flux.
The final model was chosen to be the model with the smallest $\chi^2$
measured in an area excluding pixels less than 0.65 arcsec from qA.
The improvement in the fit due to the inclusion of the S4 model was
significant ($\Delta \chi^2$ = -21).
We also fitted the profiles of the two galaxies gA and gB. For gA
the best fit was obtained with an exponential-disc profile, whereas
the best fit for gB was obtained with a de Vaucouleurs profile.

Fig.~\ref{PSFSUB}b shows a 14x14 arcsec$^2$ field of the area after
subtraction of the qA and qB PSFs as determined from the minimum
$\chi^2$ fit. The residuals, after the additional subtraction of the
final models for HGa, gA and gB can be seen directly below
(Fig.~\ref{PSFSUB}e). The magnitude of HGb was measured on this final
subtracted image.

\subsubsection{U and I band data}

The well constrained galaxy models determined from the high
signal--to--noise B
image were subsequently used to decompose the U and I--band data. Since
the combined seeing of the U--band data was poorer than that of the
B image, we first smoothed the galaxy models to the seeing of the
U--band data. For a large grid of U--band magnitudes of S4 and HGa we
then subtracted scaled versions of the smoothed S4 and HGa galaxy
models, fitted and subtracted the quasar point source component using
DAOPHOT II, and finally calculated the $\chi^2$ in an area
excluding pixels less than 0.8 arcsec from qA.
The final model was selected to be that which had the smallest
$\chi^2$. The U magnitudes of the galaxies gA and gB were determined
in the same way.
For the decomposition of the combined I--band data, which have a better
seeing than the B--band data, we first smoothed the I--band image to
the seeing of the B image and then proceeded as for the U--band data.

\begin{figure*}
\epsfig{file=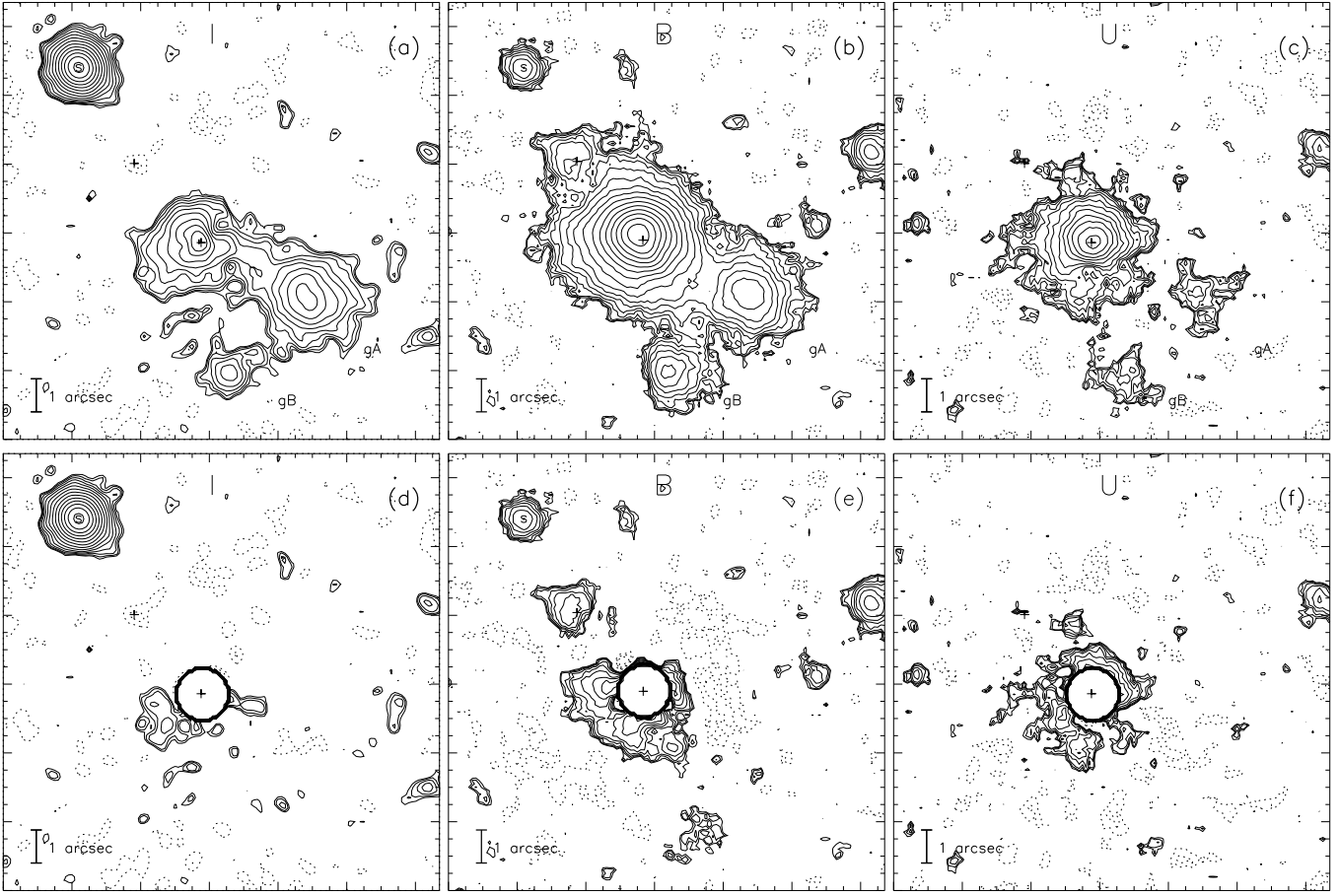,width=18cm}
\caption{Contour plots of 14x14 arcsec$^2$ surrounding the two
QSOs from the final combined and PSF subtracted frames in a) I--band, 
b) B--band and c) U--band. North is up and east to the left. Seen is residual 
extended emission under qA. Also seen are the two galaxies gA and gB south 
of qA. The contour levels are -9, -6, -3, 3 $\times$ 1$\sigma$ of the sky 
noise and thereafter spaced logarithmically in factors of 1.5, with the 
dotted contours being negative. d) -- f) show the same region after 
subtraction of the galaxy fits of HGa, gA and gB as described in the text.}
The most plausible explanation for the residual east of the position of qA 
is the absorber galaxy S4.
\label{PSFSUB}
\end{figure*}

Results of this procedure can be seen in Fig.~\ref{PSFSUB}a,d and 
Fig.~\ref{PSFSUB}c,f for the I and U--bands respectively. As for the
B--band data the upper frames show the fields after subtraction of
final fits of qA and qB only, while in the lower frames the fitted
models of galaxies HGa, gA and gB have also been subtracted.
The magnitudes (and estimated associated errors) of objects resulting
from the fitting procedure are listed in Table 3.

%=====================Begin Figure 3====================================
%=====================End Figure 3====================================

\section{Summary and discussion}

Our original interest in the field of Q0151+048A was to identify the
DLA galaxy in front of it. This identification was accomplished via
imaging in Ly$\alpha$ (Fynbo et al. 1999), but our broad band images
left some questions open. The purpose of the deeper broad--band data
presented in this paper was to clarify this situation. We shall here
first summarise our findings, then briefly consider their
implications.

\subsection{Results summary}
Our new data have unambiguously confirmed the presence of extended
emission in the field in all three bands I, B and U. The different
morphology seen in the three bands strongly suggest that we see three
objects superimposed: The quasar, the DLA absorbing galaxy and the
quasar host galaxy.

The superposition of three close objects of widely
differing brightnesses causes considerable degeneracy for any attempt
to determine the brightness of the faintest sources, and it is
therefore impossible to find a unique solution for the flux of the
faintest object (the DLA galaxy S4). Nevertheless, we find that S4 is
clearly detected in the U image. The U--band magnitude of S4
determined via our minimum $\chi^2$ procedure is fully consistent
(to within 1 $\sigma$) with being caused by the known
Ly$\alpha$ flux at 3565{\AA} alone. The data are therefore consistent
with a zero contribution from any continuum source in the U--band.

It is difficult to determine the exact errors on the I and B
magnitudes of S4, but for both images we found a very significant
improvement in the reduced $\chi^2$ of the fit when we included S4.
It is therefore likely that S4 is indeed a low surface brightness
continuum source, but this question is going to be extremely hard
to settle.

The existence of a separate extended continuum source centred on qA is,
however, clearly demonstrated independently in all bands. This result
was arrived at independently via image deconvolution, and via our
iterative object fitting technique.

\subsection{Discussion: Starburst Galaxy or Dust scattering}
The distance modulus (for z=1.93) in the assumed cosmology with
h=0.5 is 45.8. Assuming instead $\Omega$=0.3 and $\Omega_{\Lambda}=$0.7
the corresponding distance modulus becomes 46.6.
Hence, the absolute AB magnitudes of the host
galaxy HGa is $<$-24.0(-24.8) in U (rest frame 1100--1300\AA), 
$<$-24.5(-25.3) in B (rest frame 1300--1500\AA) and $<$-24.0(-24.8) 
in I (rest frame 2300--3400\AA). Such extremely bright magnitudes are
in the local universe only connected with brightest cluster galaxies
(for comparison M87 and Centaurus A both have absolute magnitudes of
roughly -23 in the V--band). Brightest cluster 
members can be as bright as -26 (Oemler 1976). Interestingly we find
that the absolute magnitude of HGa is similar to those of the
extended `fuzz' that have been detected
around other high redshift QSOs by Lehnert et al. (1992), 
Carballo et al. (1998) and Aretxaga et al. (1998a,b).

The morphology of the host galaxy HGa is best fit by a de Vaucouleurs
profile. The fit to an exponential-disc leads to a much poorer fit. 
A plausible interpretation of the data is
therefore that we see the early stage of a massive 
elliptical galaxy in the process of forming the bulk of its stars. 
Assuming that all the light is coming from stars, and not e.g.
scattered quasar light (see below), we can estimate the star formation 
rate (SFR) needed to explain the observed fluxes. In the case of
continuous star formation we can adopt the relation between the 
SFR and the luminosity at 1500\AA \  
SFR = L$_{1500}/(1.3\times10^{40} erg\: s^{-1} $\AA$^{-1}$) commonly
used for LBGs (Pettini et al. 1998) and we hence infer a star formation
rate of order 100(200) M$_{\sun}$ \ yr$^{-1}$ for $\Omega$=1(0.3) and 
$\Omega_{\Lambda}$=0(0.7). For instantaneous bursts we can use the
Starburst99 package (Leitherer et al. 1999) to infer the 
colours of models calculated with solar 
metallicity and ages 1, 10 and 100 million years. The colours 
for these three models are given in Table~\ref{instan}.

%=====================Begin Table 4====================================
\begin{table}
\begin{center}
\caption{The colours of instantaneous starbursts with four different
ages. The colours of HGa are listed for comparison.}
\begin{tabular}{@{}rrrr}
Age   & u-B  & B-I  \\
(Myr) &      &      \\
\hline
1     &  -0.2 & -0.7 \\
10    &  -0.1 & -0.3 \\
100   &   0.7 &  0.3 \\
\hline
Exta   & 0.0--1.1 & -1.1--0.1 \\
HGa   & 0.9$\pm$0.3 & -0.7$\pm$0.3 \\
\hline
\end{tabular}
\label{instan}
\end{center}
\end{table}
%=====================End Table 4====================================
The colours of the host, 0$<$u-B$<$1.1, -1.1$<$B-I$<$0.1 from 
Deconvolution and 0.9$\pm$0.3, -0.7$\pm$0.3 from PSF-subtraction, are roughly
consistent with instantaneous bursts with ages in the range 10--100
Myr. The number of stars formed in the burst would be in the range
from 10$^8$ to a few times 10$^9$ stars depending on the age of the
burst and on the assumed cosmology.

Another interpretation of the extended fuzz frequently
seen around quasars, is light from the quasar itself scattered by dust.
This mechanism is well known from radio galaxies at high redshifts where
scattering off dust grains has revealed the existence of ``hidden''
quasars in the galaxy cores. It is likely that radio quiet QSOs have
similar non--isotropic radiation fields (see e.g. the discussion
in M{\o}ller \& Kj{\ae}rgaard 1992), and in that case our line of
sight is such that we look straight down the emission cone inside of
which the scattering is taking place. In this case we therefore
expect to see the quasar emission cone ``end on'' via forward
scattered quasar light. The scattering process is expected to be
essentially grey and recent calculations predict that as much as
10\% of the quasar light could be scattered in this way (Witt \& 
Gordon, 1999; V\'arosi \& Dwek, 1999; Vernet et al. in prep.).
If considering a clumpy medium, we would expect dust scattered
light to be emitted from inside a very
large volume in front of the quasar. When taking the cone
geometry into account one would expect its total flux to be
roughly a few \% of the quasar flux at any given wavelength (Fosbury,
private communication). From Table 3 we find that the flux from HGa
is 3, 6 and 2\% of the flux from Q0151+048A in U, B and I
respectively. Similar, but less significant, results are found for
HGb. It is not yet known if the light profile of scattered 
light from a cone will reproduce a de Vaucouleurs profile, but since
this seems to be a universally preferred profile it is not unlikely.
One thing worth noting in Fig.~\ref{PSFSUB}e are the negative residuals
surrounding the position of qA at a distance of 2-3 arcsec after
subtraction of the fitted de Vaucouleurs profile. This indicates that
the true profile of HGa in reality falls off steeper than a
de Vaucouleurs profile. If model calculations were to show such a
steep profile for forward scattered light in a radiation cone, that
would be a strong hint towards the nature of the quasar fuzz.

%=====================Begin Figure 4====================================
\begin{figure}
\epsfig{file=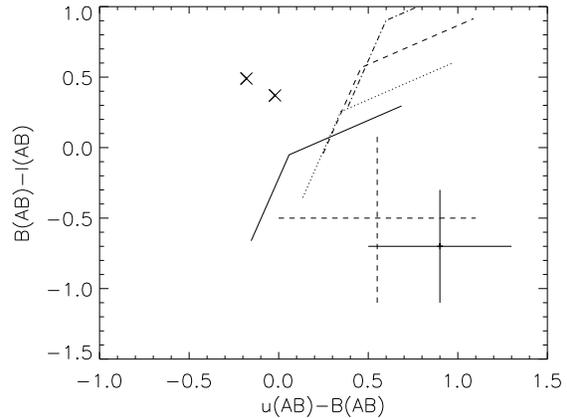,width=8cm}
\caption{The B(AB)-I(AB) vs. u(AB)-B(AB) colours of qA and qB
(marked by $\times$), of the instantaneous star burst models of 
Table 4 (thick, full drawn line). The dotted, dashed and 
long dashed lines show the colours of reddened bursts assuming 
A(B$_{Rest}$)=0.5 and MW, LMC and SMC extinction curves respectively.
The colours of the extended emission under qA is marked by the 
error-bars. The dashed error-bar represents the measurement from 
deconvolution and the full drawn error-bar the measurement from 
PSF-subtraction. 
}
\label{colours}
\end{figure}
%=====================End Figure 4====================================          

   However, the colours of the extended emission as seen in Table 3
and Fig.~\ref{colours} are significantly different from those of
the two QSOs, which argues against the scattering hypothesis. Hence,
we conclude that at least a significant fraction of the observed 
extended emission must be caused by a star burst.

\section*{Acknowledgments}
We wish to thank Pierre Magain and Peter Stetson for making available
the MCS-code and DAOHPOT-II respectively. We have benefitted from
stimulating discussions with R. Fosbury and J. Vernet on the subject
of dust scattering, and with C. Jean on the subject of reddening models 
of galaxy spectra. We thank the anonymous referee for valuable
comments that clarified the paper on essential points. 
JUF thanks the European Southern Observatory for support from the 
ESO studentship programme. IB thanks the European Southern Observatory
for support from the ESO visitor programme. IB was supported in part 
by contract ARC94/99-178 ``Action de Recherche Concert\'ee de la 
Communaut\'e Fran\c{c}aise (Belgium)'' and P\^ole d'Attraction 
Interuniversitaire, P4/05 (SSTC, Belgium).

\end{document}